\documentclass[conference,a4paper]{IEEEtran}

\ifCLASSINFOpdf
  \usepackage[pdftex]{graphicx}     \graphicspath{{./figs/}}
      \DeclareGraphicsExtensions{.pdf,.jpeg,.png}
\else
                  \fi

\usepackage[cmex10,fleqn]{amsmath}
\usepackage[fleqn]{mathtools}
\usepackage{amssymb}
\usepackage{pifont}\newcommand{\cmark}{\ding{51}}\newcommand{\xmark}{\ding{55}}

\usepackage{algorithm}
\usepackage{algpseudocode}

\usepackage{array}

\usepackage[labeled,resetlabels]{multibib}
\newcites{S,P}{Supplementary References,	Postscript References}

\usepackage{float} 
\usepackage{ulem}
\usepackage{color}
\usepackage{soul}

\usepackage{fancyhdr}

\usepackage{colortbl}
\newcommand{\nonecolor}{\cellcolor{lightgray}}  \newcommand{\nonfunc}{\cellcolor{YellowGreen}}  \newcommand{\hevc}{\cellcolor{SeaGreen}}  \newcommand{\nonfunchevc}{\cellcolor{TealBlue}}  

\usepackage{afterpage} 
\usepackage{fixltx2e} 
\usepackage{multirow}

\usepackage{randtext}
\catcode`\_=11\relax
\newcommand\email[1]{\_email #1\q_nil}
\def\_email#1@#2\q_nil{  \href{mailto:#1@#2}{{\randomize{#1}\emailampersat \randomize{#2}}}}

\newcommand\emailampersat{{\small@}} \catcode`\_=8\relax

\usepackage{doi}

\usepackage{xfrac}

\usepackage{picins} 
\usepackage{bm}
\usepackage{hyperref}
\usepackage{breakurl} \usepackage[usenames,dvipsnames]{xcolor}
\hypersetup{
    colorlinks,
    linkcolor={red!50!black},
    citecolor={blue!50!black},
    urlcolor={blue!80!black}
}

\usepackage{textcase}

\usepackage[font=small,labelfont=bf]{caption}

\usepackage[para]{footmisc}

\begingroup \makeatletter
\gdef\eqna@origamp{&} \catcode`\&\active \gdef\eqna@newamp{  \ifx\@currenvir\eqna@currenvir     \eqna@onlyfirstamp\let\eqna@onlyfirstamp\@empty   \else     \eqna@origamp   \fi
}
\gdef\eqna@hook{  \let\eqna@currenvir\@currenvir   \catcode`\&\active   \let&\eqna@newamp   \let\eqna@onlyfirstamp\eqna@origamp   }
\catcode`\*11 \gdef\eqnarray{\eqna@hook\align} \gdef\eqnarray*{\eqna@hook\align*} \global\let\endeqnarray\endalign
\global\let\endeqnarray*\endalign*
\endgroup

\usepackage{tikz}
\usetikzlibrary{calc,fit,arrows,shapes.misc}

\newcommand\markarrowtopleft[1]{    \tikz[overlay,remember picture] 
        \node (marker-#1-a) at (0,0ex) {};}
\newcommand\markarrowbottomright[1]{    \tikz[overlay,remember picture] 
        \node (marker-#1-b) at (0,0) {};	\tikz[overlay,remember picture,thick,red!100,dashed] \draw[-open triangle 45] ($(marker-#1-a.south)+(+0.07,0.06)$) -- ($(marker-#1-b.south)+(0.09,0.06)$);}

\usepackage{spreadtab}
\STautoround*{2} 
\usepackage{numprint}
\npthousandsep{,}\npthousandthpartsep{}\npdecimalsign{.}

\def\por1{\partial}

\DeclareMathOperator*{\argmax}{\,argmax}

\newcolumntype{H}{>{\setbox0=\hbox\bgroup}c<{\egroup}@{}}
\newcolumntype{M}{>{\centering\arraybackslash}m{\dimexpr0.25\linewidth-2\tabcolsep}} \newcolumntype{N}{>{\centering\arraybackslash}m{\dimexpr0.10\linewidth-2\tabcolsep}}
\newcolumntype{Y}{>{\raggedleft\arraybackslash}X}

\usepackage{caption}
\DeclareCaptionFormat{myformat}{#1#2#3\hrulefill}
\newcommand\litem[1]{\item{\bfseries #1:}}

\begin{document}
\bstctlcite{IEEEexample:BSTcontrol} 
\title{40~Gbps Access for Metro networks: Implications in terms of Sustainability and Innovation from an LCA Perspective}

\author{\IEEEauthorblockN{\small Reza \MakeTextUppercase{Farrahi Moghaddam}}
\IEEEauthorblockA{Synchromedia Lab and CIRROD}
\IEEEauthorblockA{ETS (University of Quebec)}
\IEEEauthorblockA{Montreal, QC, Canada H3C 1K3} 
\IEEEauthorblockA{Email: \email{imriss@ieee.org}} \IEEEauthorblockA{LinkedIn: \url{https://www.linkedin.com/in/rezafm}} \and
\IEEEauthorblockN{\small  Yves \MakeTextUppercase{Lemieux}}
\IEEEauthorblockA{Cloud Technology}
\IEEEauthorblockA{Ericsson Canada Inc}
\IEEEauthorblockA{Montreal, QC, Canada H4P 2N2} \and
\IEEEauthorblockN{\small  Mohamed \MakeTextUppercase{Cheriet}}
\IEEEauthorblockA{Synchromedia Lab and CIRROD}
\IEEEauthorblockA{ETS (University of Quebec)}
\IEEEauthorblockA{Montreal, QC, Canada H3C 1K3} 
}

\maketitle

\begin{abstract}
In this work, the implications of new technologies, more specifically the new optical FTTH technologies, are studied both from the functional and non-functional perspectives. In particular, some direct impacts are listed in the form of abandoning non-functional technologies, such as micro-registration, which would be implicitly required for having a functioning operation before arrival the new high-bandwidth access technologies. It is shown that such abandonment of non-functional best practices, which are mainly at the management level of ICT, immediately results in  additional consumption and environmental footprint, and also there is a chance that some other new innovations might be `missed.'  Therefore, unconstrained deployment of these access technologies is not aligned with a possible sustainable ICT picture, except if they are regulated. An approach to pricing the best practices, including both functional and non-functional technologies, is proposed in order to develop a regulation and policy framework for a sustainable broadband access.
\end{abstract}

\IEEEpeerreviewmaketitle

\begin{figure*}[tbh!]
\centering
\includegraphics[width=7in]{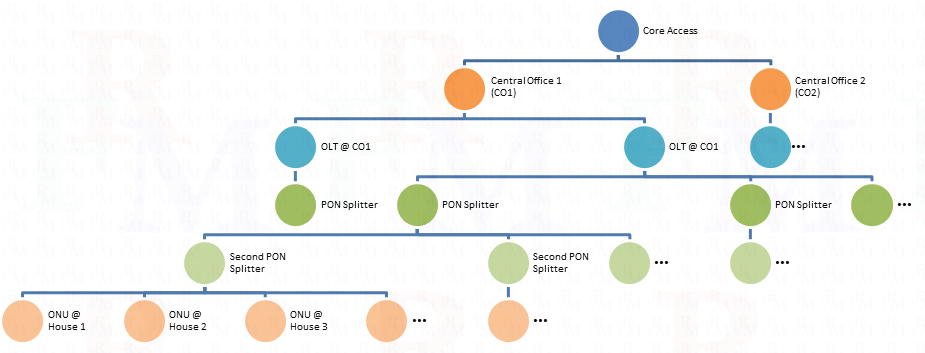}
\caption{A simplified schematic diagram of a typical access network from the metro access point to the customer premises.}
\label{fig_schematic_metro1}
\end{figure*}

\section{Introduction}
\label{sec_Introduction}
Information and Communications Technologies (ICT) has been recognized as a key sector in shifting and reducing the footprint of other sectors by providing equivalent, ICT-based services \cite{SMART2020,SMARTer2020,Farrahi2014c}. Although this so-called ICT enabling effect has been recognized as a transformer toward a more sustainable future, it has been challenged with respect to i) sustainability of ICT self-operation \cite{Arushanyan2014,Farrahi2014b}, ii) re-bound effect in terms of exponential growth in service consumption \cite{Girod2011}, and iii) implications of ICT-driven life styles \cite{Farrahi2014e}. In particular, rich digital content consumption, especially in the form of online content, has been a growing trend in both behavioral changes in the way of living and also in requirements of being considered a developed society \cite{FCC2015}. 

In terms of rich content, Online Video Services (OVSs) and Person-Recorded Videos (PRVs) are two major trends that demand a considerable connectivity bandwidth especially at the last- (first-) mile of the Internet service networks \cite{Ericsson2014,Farrahi2014e}. Other trends that require high bandwidth access networks have been smart grid, smart house, and Internet of Things trends that in their naive forms would require a considerable broadband bandwidths for telemetry connections used to pull the sensed data to central or distributed analytic sites \cite{Bates2013}. Even in the case of the mobile access, for which broadband access is planned and coined as the 5G mobile access (requiring a minimum of 50~Mbps-100~Mbps everywhere and 1~Gbps for indoor configurations by 2020),\footnote{In its current vision based on the evolution of the long term evolution-advanced and its enhancements \cite{4GAmericas2014,NGMN2015,Ericsson2015,NTTDOCOMOInc2014}} the required bandwidth could be translated as burden on the backhaul connections of the Base Transceiver Stations (BTSs) which will be probably co-tenant with fixed broadband connections.

All these new applications and requirements have made deployment and utilization of high-bandwidth broadband access network profitable (or at least justifiable in the case of municipal broadband providers). Practically, such a deployment would strongly depend on the metro access technology used \cite{Farrahi2014f}. Many new metro access technologies, mainly populated around Passive Optical Networks (PONs), have been explored and deployed \cite{Lambert2014} (for a simplified schematic diagram, please see Figure \ref{fig_schematic_metro1}). Although these technologies would improve the quality of life and the level of development, they could have a side effect in terms of `disabling' sustainability mechanisms that were before applicable because of low amount of available resources, i.e., the broadband bandwidth. These mechanisms, such as the micro-registration mechanism \cite{Farrahi2014f}, could be usually considered to be non-functional because they would become transparent when the level of available resource excesses a specific threshold. 

In this work, we evaluate the impact of high-bandwidth access technologies on making non-functional technologies inapplicable, and then we try to implement smart pricing mechanisms that not only bring the non-functional mechanisms and technologies back to the picture, they also `encourage' the operators and the holders of the best practices to implement the best practices toward a sustainable operation.

\section{Functional and Non-Functional Technologies}
\label{sec_Functional_NonFunctional_Technologies}
In any service operation, there are parts that are essential to the operation. For example, in the Internet access service provided by the ISPs, actual connection (wired or wireless) is an obvious essential part of the service operation. These parts of the operation could be located at various layers in a layer-decomposition model of the operation, for example at the infrastructure layer. At the same time they could appear in various forms, such as hardware, software components, and applications. It is common to consider these essential parts of the operation as the `functional' parts. In contrast, those parts that are not essential, i.e., they have been added as extra `features,' would be considered as the `non-functional' parts of the operation. In this traditional picture, adding, dropping, or changing the grade of a non-functional part would not violate the original agreement between the parties involved in a service operation. As a consequence, a service user could change the non-functional features of its service simply by issuing `on-the-fly' agreements to amend their original contract of service. 

To be more generic in this work, we only focus on the technologies themselves. Any part of an operation is rooted in an associated technology. Therefore, by analyzing the technologies, the associated parts at any layer would be implicitly covered in a generic analysis. In particular, we are interested in the metro access network, and therefore the related technologies are of our great interest. However, as will be seen in the following sections, some of technologies related to the last- [first-] mile access, such as the micro-registration technology in Section \ref{sec_Practices_Best_Practices_NonFunctional_Technologies_Access_Network}, would have footprint outside the metro access area, and would even reach the Content Delivery networks (CDNs). 

The first goal of this work is to illustrate that the functional/non-functional devision of technologies is no longer a solid line, and many previously-thought non-functional technologies could actually play the role of an enabling functional technologies when properly combined with other technologies. The second objective is to study the phenomenon of neglecting such enabling technologies when the new core functional technologies could deliver the service without requiring the implementation of the non-functional technologies. In this case, although the resulting service operation would be functional, its energy consumption and other environmental impacts would not be minimal. Therefore, we will propose a pricing model in order to enforce implementation of the best practices.

\section{Practices and Best Practices of Last-Mile Access Network Technologies}
\label{sec_Practices_Best_Practices_Edge_Technologies_Access_Network}
To be more accurate, we assume a specific configuration in the access network. Following \cite{Lambert2014}, it is assumed that the access network is a Two-level Fiber-to-the-Home (FTTH)\footnote{Or, equivalently, Fiber-to-the-Point (FTTP).} in the form of a Passive-Optical-Network (PON). This means that there are two levels ($N_l=2$) that are implemented using splitters.\footnote{This is not the case for the Copper-to-the-Home access networks, i.e., the OC-48 access technology below. However, we skip the details of that case for the purpose of simplicity.} Although there are some recommendations in terms of minimum bandwidth required for each home, such as those of FCC,\footnote{FCC's requires a broadband access of 25Mbps/3Mbps downstream/upstream (DS/US) for fixed access, and of 10Mbps/768kbps DS/US for mobile access \cite{FCC2015}.} we will consider a set of scenarios in order to assess the performance of various technologies. These scenarios are provided in section \ref{sec_broadband_scenarios}.

In terms of reach of an optical technology, the Optical Budget (OB) and attenuation ($\alpha$) of that technology and also the split ratio at each level $l$ ($S_l$) would determine the maximum reachable distance:
\begin{eqnarray}
d_\text{max} & \leq & \frac{1}{\alpha} \left(\text{OB} - L_m - \sum_{l=1}^{N_l} 3.5 \times \log_2S_l\right) \\
& = & \frac{1}{\alpha} \left(\text{OB} - L_m - 3.5 \log_2S\right)
\end{eqnarray}
where $L_m = 3$dB is a margin for fiber patching \cite{Lambert2014}, $S=\Pi_{l=1}^{N_l} S_l$ is the total split ratio, and $\mathbf{S}=\left(S_1, \cdots, S_l\right)$.

\begin{table*}
\centering
\setlength{\tabcolsep}{2pt}
\begin{tabular}{||c|l||c|c|HH|c|c|c|c||}\hline\hline
Label & Technology & \multicolumn{8}{c||}{Features} \\\cline{3-10}
 &  & \begin{tabular}{c}D BW\\(Gbps)\end{tabular} & \begin{tabular}{c}U BW\\(Gbps)\end{tabular} & \begin{tabular}{c}Optical\\ Budget, OB (dB)\end{tabular} & \begin{tabular}{c}Attenuation\\ $\alpha$ (dB/km)\end{tabular} & \begin{tabular}{c}$d_\text{max}$ (km)\\ $@\mathbf{S}=\left(8, 8\right)$\end{tabular} & \begin{tabular}{c}$d_\text{max}$ (km)\\ $@\mathbf{S}=\left(8, 16\right)$\end{tabular} &
 \begin{tabular}{c}$d_\text{max}$ (km)\\ $@\mathbf{S}=\left(16, 16\right)$\end{tabular} &
 \begin{tabular}{c}$d_\text{max}$ (km)\\ $@\mathbf{S}=\left(32, 16\right)$\end{tabular} \\\hline\hline
Ta & OC-48 & 2.49$^\dagger$ & 2.49 & N/A & N/A & 4.5$^\ddagger$ & 4.5$^\ddagger$  & 4.5$^\ddagger$ & N/A \\\hline
Tb & GPON B+ & 2.50 & 1.25 & 28.0 & 0.6 & 6.7 & 0.8 & N/A & N/A \\\hline
\multirow{2}{*}{Tc} & 40G TDM PON:GEM & \multirow{2}{*}{40.00} & \multirow{2}{*}{10.00} & \multirow{2}{*}{31.0} & \multirow{2}{*}{0.6} &  \multirow{2}{*}{11.7} & \multirow{2}{*}{5.8} & \multirow{2}{*}{N/A} & \multirow{2}{*}{N/A} \\
 & 40G TDM PON:BI & & & & & & & & \multicolumn{1}{c||}{}  \\\hline
Td & TWDM & 40.00 & 10.00 & 35.0 & 0.4 & 27.50 & 18.8 & 10.0 & 1.3\\\hline
Te & OFDM & 40.00 & 10.00 & 34.5 & 0.6 & 17.50 & 11.7 & 0.6 & N/A\\\hline\hline
\end{tabular}
\caption{The specifications of various metro access technologies \cite{Lambert2014,Farrahi2014f}. D BW and U BW stands for download and upload bandwidths. Maximum reach distance for various split ratios are provided. Notes: $^\dagger$The overhead bandwidth is ignored. $^\ddagger$7/3~Mbps DS/US using Asymmetric-bit-rate Digital Subscriber Lines (ADSL) or Very-high-bit-rate DSL (VDSL) \cite{Kartalopoulos2004}.}
\label{tab_PON_Tech1}
\end{table*}

A list of the PON access technologies and their specific parameters are provided in Table \ref{tab_PON_Tech1}.\footnote{The acronyms: Optical Carrier-48 (OC-48) \cite{Izadpanah1992}, Gigabit-Capable Passive Optical Networks (GPON) \cite{ITUTG984x}, 10-gigabit-capable passive optical networks (XG-PON) \cite{ITUTG987x}, Ethernet Passive Optical Network (EPON) and 10/1G-EPON \cite{IEEE802.3}, GPON Encapsulation Method (GEM) \cite{Cale2007}, Time-and Wavelength-Division Multiplexed (TWDM) \cite{Williams1993}, and optical Orthogonal Frequency Division Multiplexing (OFDM) \cite{Cvijetic2012}.} In short, a copper-based technology, a low-bandwidth (2.5~Gbps) optical technology, and three high-bandwidth (40~Gbps) optical technologies are considered.
It is worth mentioning that the access technologies are not limited to those listed in Table \ref{tab_PON_Tech1}. For example, a split ratio of 8192 splits over a distance of 135km with symmetric capacity of 320~Gbps using DWDM-TDMA PON  was reported in \cite{Ossieur2011}.\footnote{DWDM-TDMA stands for Dense Wavelength-Division-Multiplexed Time-Division Multiple Access \cite{Ossieur2011}.} A 10~Gbps TDM-OCDM-PON system without an en/decoder at the optical network
unit at the customer premises was reported in \cite{Kodama2013}.\footnote{TDM-OCDM: Time Division Multiplexing Optical Code
Division Multiplexing \cite{Kodama2013}.} In hybrid approaches, a combination of the existing technologies as reported in \cite{Matrakidis2015} could be considered. This would enable long range last-``mile'' access that goes beyond hundred of kilometers and make the metro access network (MAN) pointless even for rural areas. And, finally, quantum data transfer  \cite{Weiler2015} could be mentioned. However, as mentioned before, the purpose of this work is not listing all technologies. Instead, we are looking to show the implications of a few new technologies, and then ways to make them sustainability-friendly.

\subsection{Video Encoding Technologies}
\label{sec_Video_Encoding_Technologies}
In terms of online video services, the video encoding plays a critical role, and this seems to be a great potential for deployment of new services and also reduction of the associated transport-related footprint. Usually, a video content is horizontally classified in terms of its resolution in the form of two popular classes: i) HD video (1080 pixels) and ii) 4K video (2016 pixels). Interestingly, the distance beyond that the pixels cannot be perceived (a lower distance is better), for a fixed TV size such as 50-inch TV drastically reduces from 2.2 meter for HD contents to 1.0 meter for 4K contents \cite{Yardley2014}.
Along a vertical dimension, the video encoding is also divided into two groups:
\begin{LaTeXenumerate}
\litem{Advanced Video Coding (AVC)} The current approved encoder under AVC group is MPEG-4 Part 10 AVC (in short, H.264/AVC), which is approved 2003 \cite{Caron2013}.\\
\litem{High Efficiency Video Coding (HEVC)} Mostly known as H.265, its current approved version is MPEG-H, HEVC, Part 2 (in short, H.265/HEVC), which approved 2013 \cite{Grois2013,Grois2014,Diniz2015}.
\end{LaTeXenumerate}
It has been observed that H.264 has 40.8\% bit rate overhead (in terms of Bj{\o}ntegaard-Delta bit-rate (BD-BR) measure \cite{Bjontegaarda2001,Bjontegaard2008}) compared to H.265 \cite{Grois2014,Rerabek2014}.
In particular, it has been reported that the worst performance of HEVC almost collides with the best performance of the AVC for both HD and 4K video content  \cite{Yardley2014} (see Table \ref{tab_AVC_vs_HEVC_1}).
\begin{table}[!htb]
\centering
\begin{tabular}{||l||c|c||c||}\hline\hline
Video & AVC low & HEVC high & HEVC low \\\hline\hline
HD & 6~Mbps & 4.9~Mbps & 3.0~Mbps\\\hline
4K & 16~Mbps & 20~Mbps & 8.0~Mbps\\\hline\hline
\end{tabular}
\caption{The performance of H.264/AVC and H.265/HEVC in terms of required broadband bandwidth to deliver HD and 4K videos \cite{Yardley2014}.}
\label{tab_AVC_vs_HEVC_1}
\end{table}
Therefore, there is a big potential for the HEVC and its succeeding encoding approaches to relive the ongoing increase in demand for bandwidth related to OVSs. In the scenarios considered in this work, we use the lower-bound (best performance) of the video encoders.

\section{Practices and Best Practices of `Non-Functional' Access Network Technologies}
\label{sec_Practices_Best_Practices_NonFunctional_Technologies_Access_Network}
We would prefer to refer the approaches and technologies listed in this section in the form of non-functional technologies in that sense that they are not essential to provide a broadband access service. However, as will be seen in section \ref{sec_100_vs_2_acess}, for some combinations of scenarios and access technologies, the service would be impossible without considering one of these non-functional technologies. 

In general, the non-functional technologies seem to be positioned at the management layer. This is of great interest because the management layer, and more generally software, is the most flexible part of the ICT. Therefore, there is a good opportunity to `upgrade' and improve already deployed systems by modifying the software. However, it is worth mentioning that high-degree of dependency among software components could lead to increased inertia and 
regidity of software \cite{Farrahi2014c}. 

We list three possible technologies in this section considering the broadband application. However, it should be mentioned that the options are not limited to this list, and we use these examples only for the purpose of illustrating the concept. To be specific, the technologies are:
\begin{LaTeXenumerate}
\litem{Micro-registration} This approach, and its generalization in the form of Delayed Micro-registration, were introduced as possible answers to the Over-the-Top (OTT) Online-Video-Service (OVS) broadband bandwidth bottleneck especially in evening prime time \cite{Farrahi2014f,Farrahi2014e}. In contrast to TV and IPTV cases, Subscribed Video-on-Demand (SVoD) services could not immediately leverage on aggregation of video streams, which is a critical factor in bandwidth limited MANs. Micro-registration approaches leverage on the small but significant adaptability of viewers toward initiating a form of synchronization at short intervals of time (for example, 5~seconds) among the video streams. It has been observed such an action could reduce the number of active streams from 5,000 streams to 360 streams in some use cases that would not only increase the quality of experience of the users despite having scarce resources, it also would result in less footprint because less number of video segment packets would be required to be transmitted. To be accurate, the 360 HD streams at a bandwidth of 5~Mbps would consume a total bandwidth of 1.76~Gbps compared to flooded full bandwidth of 2.49~Gbps (considering OC-48 as the metro access technology). This translates into 29\% reduction in bandwidth requirement and its associated energy consumption and environmental footprint achieved without any modification to configuration or upgrade in hardware.\\
\litem{Edge caching} In this technique, the video segment files are cached in closest routing units to the end user, and requests from other users that match the same stream are served locally without transferring the request to the associated Content Distribution Network (CDN) server \cite{Broadbent2012,Kuenzer2013,StreamingVideoAlliance2014,Broadbent2014,Georgopoulos2014,Broadbent2015}. Considering the fact that currently most of the CDN equipment is placed before the metro access network, this caching technique would help to reduce the bandwidth demand at the metro bottleneck. \\
\litem{ Local P2P content sharing at the last mile} Point-to-Point (P2P) approaches to distribution of data and media have been reached to certain level of maturity that the idea of using such techniques to implement some sort of distributed penetration of the CDNs beyond the metro access seems feasible and practical \cite{Salvador2014}. In particular, encryption and containerization of media has been considered along with P2P approaches in order to enforce the media rights even when the media carrier and host is located outside the boundaries of the media right owner \cite{Cieply2014}.
\end{LaTeXenumerate}

\section{Broadband Scenarios}
\label{sec_broadband_scenarios}
In this section, a series of possible broadband access scenarios are listed to be used in the comparison of performance of the functional and non-functional access technologies:
\begin{LaTeXenumerate}
\litem{Sc1 --- Prime time with 1 OTT Video} In this scenario, only the prime time in the evening is considered. It is assumed that there is a single popular video (probably an episode of a series) that is being watched simultaneously but in an asynchronous way by a large portion of a metro area. To be consistent with \cite{Farrahi2014f}, it is assumed that the HD content requires a broadband bandwidth of 5.26~Mbps (Please see Section \ref{sec_Video_Encoding_Technologies} for more discussion of video encoding technologies).\footnote{It is worth noting that in the comparison made in Section \ref{sec_100_vs_2_acess} we use the values provided in Section \ref{sec_Video_Encoding_Technologies} instead of 5.26~Mbps for the AVC encoding in order to have a more consistent comparison with the HEVC encoding.} Also, the prime time is limited to the time period from 6PM to 8PM having all the end users send their requests within the first 30 minutes of this period.  If the estimated population of the metro area is around 1,000,\footnote{In contrast to the EU, in Canada, and North America in general, providing access and broadband to rural areas is a challenge and at the same time a burden in front of further investment in the urban areas.} the raw DS required bandwidth at an asynchronous state would be 5.14~Gbps.\footnote{It is worth mentioning that we consider aggregation techniques such as statistical multiplexing \cite{Pulipaka2013} under the non-functional technologies, in particular in the form of distributed, short-interval caching techniques. In addition, in many OVSs, video file segments are the smallest unit of transmission, and therefore video frames are not immediately available for the frame-based aggregation techniques.}
\litem{Sc2 --- VoIP + IPTV + Internet} This scenario corresponds to the case considered in \cite{Lambert2014}. This would require a dedicated 9.73~Mbps ($=5.26~\text{Mbps}*1.85$) per home for the IPTV service (assuming 1.85 TV channels per home, and also smart channel zapping \cite{Lee2010a,Ramos2013a}). An average, shared bandwidth of 1~Mbps per user is assumed for the Internet service in the best-effort way. The (sub-)metro area is considered to be associated to a split ratio of $S=256$. The total required bandwidth would be 2.68~Gbps.
\litem{Sc3 --- Extended Prime Time} In recent analyses, it has been observed that the video watching behavior has been evolved over years, and many age demographic groups have diverged from each other not only in terms of video content but also in terms of device used for video watching and also actual time period of 'prime time' \cite{Ericsson2014}. This scenario could also reflect the move of cable-cutting in its hard (actual dropping the subscription) or soft (transformation of cable companies into OTT Internet service and OVS providers \cite{Hamzeh2015}) forms. In either cases, IPTV would be almost meaningless. Again, a split ratio of $S=256$ is assumed. In contrast to the scenario Sc2, there will be no reserved 1~Mbps bandwidth for the Internet. 
\litem{Sc4 --- 4K Content} With availability of 4K video contents on the VoD services and also popularity of the compatible TVs, in this scenario an upgraded version of the scenario Sc3 is considered in which 4K OTT video content is considered. Both $S=256$ and having 1.85 channels watched simultaneously assumptions are shared among the scenarios Sc2, Sc3, and Sc4.
\end{LaTeXenumerate}
Considering the reach distances provided in Table \ref{tab_PON_Tech1} and the split ratio of the scenarios, the access technologies Tb and Tc would not be applicable in any scenario. Therefore, we use a split ratio of $S=128$ for these two technologies in the scenarios Sc2, Sc3, and Sc4.

\section{When Non-Functional Becomes Functional}
\label{sec_100_vs_2_acess}
In this section, the mesh of functional and non-functional access technologies against the broadband scenarios and video encoding is presented. To be concise, the results are provided in the form of a operability table, Table \ref{tab_scenario_vs_technology_1}.

\begin{table*}[!htb]
\tiny
\centering
\setlength{\tabcolsep}{4pt}
\begin{tabular}{||l|l||c|c||c|c||c|c||c|c||c|c||}\hline\hline
\multicolumn{2}{||c||}{} & \multicolumn{10}{c||}{Technologies}\\\hline\hline
Scenario & Encoding & \multicolumn{2}{c||}{Functional Ta} & \multicolumn{2}{c||}{Functional Tb} & \multicolumn{2}{c||}{Functional Tc} & \multicolumn{2}{c||}{Functional Td} & \multicolumn{2}{c||}{Functional Te} \\\hline
\multicolumn{2}{||c||}{} & \begin{tabular}{c} w/o\\ non-func.\end{tabular} & \begin{tabular}{c} w/\\ non-func.\end{tabular}  & \begin{tabular}{c} w/o\\ non-func.\end{tabular} & \begin{tabular}{c} w/\\ non-func.\end{tabular}  & \begin{tabular}{c} w/o\\ non-func.\end{tabular} & \begin{tabular}{c} w/\\ non-func.\end{tabular}  & \begin{tabular}{c} w/o\\ non-func.\end{tabular} & \begin{tabular}{c} w/\\ non-func.\end{tabular}  & \begin{tabular}{c} w/o\\ non-func.\end{tabular} & \begin{tabular}{c} w/\\ non-func.\end{tabular}  \\\hline\hline
\multirow{2}{*}{Sc1} & AVC & 
\nonecolor\textcolor{BrickRed}{\xmark} & \nonfunc\textcolor{ForestGreen}{\cmark}&
\nonecolor\textcolor{BrickRed}{\xmark} & \nonfunc\textcolor{BrickRed}{\xmark}&
\nonecolor\textcolor{BrickRed}{\xmark} & \nonfunc\textcolor{BrickRed}{\xmark}&
\nonecolor\textcolor{BrickRed}{\xmark} & \nonfunc\textcolor{BrickRed}{\xmark}&
\nonecolor\textcolor{BrickRed}{\xmark} & \nonfunc\textcolor{BrickRed}{\xmark} \\\cline{2-12}
								 & HEVC & 
\hevc\textcolor{BrickRed}{\xmark} &  \nonfunchevc\textcolor{ForestGreen}{\cmark}&
\hevc\textcolor{BrickRed}{\xmark} &  \nonfunchevc\textcolor{BrickRed}{\xmark}&
\hevc\textcolor{BrickRed}{\xmark} &  \nonfunchevc\textcolor{BrickRed}{\xmark}&
\hevc\textcolor{BrickRed}{\xmark} & \nonfunchevc\textcolor{BrickRed}{\xmark}&
\hevc\textcolor{BrickRed}{\xmark} &  \nonfunchevc\textcolor{BrickRed}{\xmark} \\\hline\hline

\multirow{2}{*}{Sc2} & AVC & 
\nonecolor\textcolor{BrickRed}{\xmark} & \nonfunc\textcolor{BrickRed}{\xmark}&
\nonecolor\textcolor{BrickRed}{\xmark} & \nonfunc\textcolor{BrickRed}{\xmark}&
\nonecolor\textcolor{ForestGreen}{\cmark} & \nonfunc\textcolor{ForestGreen}{\cmark} &
\nonecolor\textcolor{ForestGreen}{\cmark} & \nonfunc\textcolor{ForestGreen}{\cmark}&
\nonecolor\textcolor{ForestGreen}{\cmark} & \nonfunc\textcolor{ForestGreen}{\cmark} \\\cline{2-12}
								 & HEVC & 
\hevc\textcolor{ForestGreen}{\cmark} &  \nonfunchevc\textcolor{ForestGreen}{\cmark} &
\hevc\textcolor{ForestGreen}{\cmark} &  \nonfunchevc\textcolor{ForestGreen}{\cmark} &
\hevc\textcolor{ForestGreen}{\cmark} &  \nonfunchevc\textcolor{ForestGreen}{\cmark} &
\hevc\textcolor{ForestGreen}{\cmark} &  \nonfunchevc\textcolor{ForestGreen}{\cmark} &
\hevc\textcolor{ForestGreen}{\cmark} &  \nonfunchevc\textcolor{ForestGreen}{\cmark} \\\hline\hline

\multirow{2}{*}{Sc3} & AVC & 
\nonecolor\textcolor{BrickRed}{\xmark} & \nonfunc\textcolor{BrickRed}{\xmark}&
\nonecolor\textcolor{BrickRed}{\xmark} & \nonfunc\textcolor{ForestGreen}{\cmark}&
\nonecolor\textcolor{ForestGreen}{\cmark} & \nonfunc\textcolor{ForestGreen}{\cmark} &
\nonecolor\textcolor{ForestGreen}{\cmark} & \nonfunc\textcolor{ForestGreen}{\cmark}&
\nonecolor\textcolor{ForestGreen}{\cmark} & \nonfunc\textcolor{ForestGreen}{\cmark} \\\cline{2-12}
								 & HEVC & 
\hevc\textcolor{ForestGreen}{\cmark} &  \nonfunchevc\textcolor{ForestGreen}{\cmark}&
\hevc\textcolor{ForestGreen}{\cmark} &  \nonfunchevc\textcolor{ForestGreen}{\cmark}&
\hevc\textcolor{ForestGreen}{\cmark} & \nonfunchevc\textcolor{ForestGreen}{\cmark} &
\hevc\textcolor{ForestGreen}{\cmark} & \nonfunchevc\textcolor{ForestGreen}{\cmark}&
\hevc\textcolor{ForestGreen}{\cmark} & \nonfunchevc\textcolor{ForestGreen}{\cmark}
\\\hline\hline

\multirow{2}{*}{Sc4} & AVC & 
\nonecolor\textcolor{BrickRed}{\xmark} & \nonfunc\textcolor{BrickRed}{\xmark}&
\nonecolor\textcolor{BrickRed}{\xmark} & \nonfunc\textcolor{BrickRed}{\xmark}&
\nonecolor\textcolor{ForestGreen}{\cmark} & \nonfunc\textcolor{ForestGreen}{\cmark} &
\nonecolor\textcolor{ForestGreen}{\cmark} & \nonfunc\textcolor{ForestGreen}{\cmark}&
\nonecolor\textcolor{ForestGreen}{\cmark} & \nonfunc\textcolor{ForestGreen}{\cmark} \\\cline{2-12}
								 & HEVC & 
\hevc\textcolor{BrickRed}{\xmark} &  \nonfunchevc\textcolor{BrickRed}{\xmark}&
\hevc\textcolor{BrickRed}{\xmark} &  \nonfunchevc\textcolor{ForestGreen}{\cmark}&
\hevc\textcolor{ForestGreen}{\cmark} & \nonfunchevc\textcolor{ForestGreen}{\cmark} &
\hevc\textcolor{ForestGreen}{\cmark} &  \nonfunchevc\textcolor{ForestGreen}{\cmark}&
\hevc\textcolor{ForestGreen}{\cmark} &  \nonfunchevc\textcolor{ForestGreen}{\cmark} \\\hline\hline
\end{tabular}
\caption{A mesh of combinations of access technologies, non-functional technologies, broadband scenarios, and video encoding technologies. The feasible combinations are marked with \textcolor{ForestGreen}{\cmark}, while the rest are highlighted with \textcolor{BrickRed}{\xmark}. For the technologies Tb and Tc, a split ratio of $S=128$ is considered.}
\label{tab_scenario_vs_technology_1}
\end{table*}

As can be seen from the table, the non-functional technologies and also HEVC have a major contribution in making access technologies functional when the access technology's bandwidth is less than 40~Gbps. In particular, the summary of the contribution of these two enhancements in presented in Table \ref{tab_scenario_vs_technology_sum1}. The table compares the number of access technology and scenario pairs that become functional using one or both of enhancements. Interestingly, access technologies with less than 40~Gbps access are not functional in all 8 cases if no enhancement is considered. When both non-functional technologies and HEVC enhancements are considered, 6 out of 8 cases become functional. In contrary, for access technologies with a bandwidth higher than 40~Gbps, enhancements do not provide any improvement, and the number of functional cases stays 6 out 12 for all combinations. This conclusion would be simply translated in discarding or neglecting the `impacts' of the enhancements because their functionality-related impact, which is the main impact directly influencing the business, is NIL. In the next section, we will see that the impact on the environmental footprint and resource consumption, i.e., the energy consumption, is not NIL, and therefore a smart intervention is required to make the enhancements meaningful. We will consider a resource consumtion pricing approach in the following sections to address this requirement.

\begin{table}[!htb]
\centering
\setlength{\tabcolsep}{2pt}
\begin{tabular}{||M|M||c|c||}\hline\hline
\multicolumn{2}{||c||}{Enhancement} & \begin{tabular}{c}
$<40$~Gbps\\ Technologies\end{tabular} & \begin{tabular}{c} $>40$~Gbps\\ Technologies\end{tabular} \\\hline\hline
\nonecolor$\varnothing$\markarrowtopleft{a1} & \markarrowbottomright{a1}\nonecolor$\varnothing$ & 0/8 & 9/12 \\\hline
\nonecolor$\varnothing$\markarrowtopleft{a2} & \markarrowbottomright{a2}\nonfunc Non-func. & 4/0 & 0/9 \\\hline
\nonecolor$\varnothing$\markarrowtopleft{a3} & \markarrowbottomright{a3}\hevc HEVC & 4/0 & 0/9 \\\hline
\nonecolor$\varnothing$\markarrowtopleft{a4} &  \markarrowbottomright{a4}\nonfunchevc Non-func. $\oplus$ HEVC & 6/0 & 0/9 \\\hline\hline
\end{tabular}
\caption{The summary of Table \ref{tab_scenario_vs_technology_1} in terms of number of cases the enhancements have made a combination functional.}
\label{tab_scenario_vs_technology_sum1}
\end{table}

\section{An LCA footprint Comparison of Combinations of Access Technologies and Improvement Practices}
\label{sec_LCA_footprint_Comparison_100_and_2_Cases}
For each tuple of access technologies, scenarios, and enhancements, a full analysis of energy consumption would be required to have an LCA.\footnote{Application of LCA for ICT has various challenges that are mostly exclusive to this industry. Please see \ref{sec_Challenges_Adaptation_LCA_ICT} for a short, selected list.} However, many components are shared and would not change when moving from one tuple to another one. Therefore, a differential analysis could be considered. Also, considering the unavailability of the manufacturing phase's energy consumption and footprint data specific for each technology and also similarity among the technologies in terms of volume and wight of equipment, the embodied consumption and footprint is ignored. That said, it is worth mentioning that some parts of the CPE are technology-dependent, and therefore they should be accounted for in the differential analysis. To be more specific, the optical network unit (ONU) part of the CPE shows dependencies on both technology and also bandwidth \cite{Lambert2014}. The total differential electricity consumption per active home could be summarized as follows:

\begin{eqnarray}
P_\Delta & = & P_{\Delta \text{OLT}} + P_{\Delta \text{ONU}} \nonumber \\
& = & \frac{1}{N_h} \left(P_{\text{OLT}, \text{port}} + N_h P_{\text{OLT}, \text{user}}\right) + P_{\Delta \text{ONU}} \nonumber \\
& = & \frac{1}{N_h} \left(P_{\text{OLT}, \text{port}} + N_h P_{\text{OLT}, \text{user}}\right) + \nonumber \\
& & \left(P_{\text{ONU},0} \frac{N_s}{N_h} + P_{\text{ONU},1}  \text{BW}_{\text{D}}\right) \nonumber \\
& = & \frac{1}{N_h} \left(P_{\text{OLT}, \text{port}} + N_h P_{\text{OLT}, \text{user}}\right) + \nonumber \\
& &  \left\{P_{\text{ONU},00} + \frac{1~\text{Gbps} *1~\text{W}}{1~\text{Gbps} - 100~\text{Mbps}} + \right. \nonumber \\ 
& & \left. \frac{-100~\text{Mbps} *1~\text{W}}{1~\text{Gbps} - 100~\text{Mbps}}\right\} \frac{N_s}{N_h}  +  \nonumber \\
& & \left(\frac{N_{h0} P_{\Delta \text{OLT}0} \text{BW}_{\text{D}}}{1~\text{Gbps} - 100~\text{Mbps}}\right)\left(\frac{1}{40} - \frac{1}{N_{s,0}}\right)
\label{eq_power_per_user_1}
\end{eqnarray}
where $P_{\text{OLT}, \text{port}}$ is the power required per port of an optical line terminal (OLT) at a Central Office (CO), $P_{\text{OLT}, \text{user}}$ is the power per connected home, $N_h$ is the number of active homes at the bandwidth $\text{BW}_{\text{D}}$, and $P_{\Delta \text{ONU}, 0}$ is the zero-offset power required by an ONU at a home at 1~Gbps, and $N_{s,0}$ is the split number that provides both 100~Mbps and 1~Gbps \cite{Lambert2014}. The $N_{s,0}$ and $P_{\text{ONU},00}$ values are 64, 64, 256, and 128 and 8~W, 13~W, 12~W, and 19~W respectively for Tb, Tc, Td, and Te access technologies of Table \ref{tab_PON_Tech1} \cite{Lambert2014}. The parameter $N_h$ can be simply calculated as follows:
\begin{equation}
N_h = \min(N_s, \text{BW}_\text{max} / \text{BW}_\text{D}),
\end{equation}
where $\text{BW}_\text{max}$ is the maximum DS bandwidth of an access technology. 
Also, $N_{h0}$ and $P_{\Delta \text{OLT}0}$ are those values of $N_h$ and $P_{\Delta \text{OLT}}$ that are calculated at 100~Mbps bandwidth. Using Equation \ref{eq_power_per_user_1}, the energy consumption per Gb of download, which takes $1 \text{Gb}/\text{BW}_{\text{D}}$ seconds at $\text{BW}_{\text{D}}$ bandwidth, will be:
\begin{eqnarray}
E_\Delta & = & \frac{1~\text{Gb}}{\text{BW}_{\text{D}}}\left(P_{\Delta \text{OLT}} + P_{\Delta \text{ONU}}\right) \nonumber \\
& = & \frac{1~\text{Gb}}{N_h \text{BW}_{\text{D}}} \left(P_{\text{OLT}, \text{port}} + N_h P_{\text{OLT}, \text{user}}\right) + \nonumber \\
& &  1~\text{W}*\left\{\frac{1~\text{Gb}/1~\text{W}}{\text{BW}_{\text{D}}}P_{\text{ONU},00} + \frac{1~\text{Gb}/\text{BW}_{\text{D}}}{1 - 100/1024} + \right. \nonumber \\ 
& & \left. \frac{-100~\text{Mb}/\text{BW}_{\text{D}}}{1 - 100/1024}\right\} \frac{N_s}{N_h}  +  \nonumber \\
& & \left(\frac{N_{h0} P_{\Delta \text{OLT}0}* 1~\text{s}}{1 - 100/1024}\right)\left(\frac{1}{40} - \frac{1}{N_{s,0}}\right) 
\label{eq_power_per_MB_1}
\end{eqnarray}
Or in a short form:
\begin{equation}
E_\Delta  = \frac{A_\Delta}{\text{BW}_{\text{D}}} + B_\Delta
\label{eq_power_per_MB_2}
\end{equation}
where $A_\Delta$ and $B_\Delta$ are technology-dependent coefficient. The calculated values of these coefficients for various technologies is provided in Table \ref{tab_tech_energy_coeff_1} ($N_s=256$ align with the scenarios of Section \ref{sec_broadband_scenarios}).

\begin{table}[!htb]
\centering
\begin{tabular}{||l||c|c||}\hline\hline
Technology & $A_\Delta$ (WMb) & $B_\Delta$ (J) \\\hline\hline
Tb & 9,228.0 & 0.0312 \\\hline
Tc & 14,480.0 & 0.3751 \\\hline
Td & 13,531.0 & 1.2810 \\\hline
Te & 21,368.0 & 4.2286 \\\hline\hline
\end{tabular}
\caption{The calculated coefficients of Equation \ref{eq_power_per_MB_2} for various technologies. The data provided in \cite{Lambert2014} is used for this calculations. $N_s=256$ and $\text{BW}_{\text{D}}=6.55~\text{Mbps}$.}
\label{tab_tech_energy_coeff_1}
\end{table}

The effective energy consumption per 1~Gb of download is provided in Table \ref{tab_scenario_vs_technology_LCA1}. For the technology Ta, the average values of metro access energy consumption could be used \cite{Malmodin2014}. However, we do not consider it because this technology is highly different from the others, and therefore other factors could influence the comparison. It can be seen from the table that the access technology Tb outperforms the 40~Gbps technologies but only when it is combined with both non-functional technologies and the HEVC video encoding. Even for high bandwidth access technologies, the energy consumption is almost half when the enhancements are considered.\footnote{It is worth mentioning that the energy consumption per 1 second of HD video watched is almost the same when the non-functional technologies of Section \ref{sec_Practices_Best_Practices_NonFunctional_Technologies_Access_Network} are not considered and only AVC and HEVC video encodings are compared: For the case of the AVC video encoding, and for {\em (functional Tb, w/o non-functional technologies, Sc3, HD, AVC)} tuple of technology combination, the energy consumption per 1 second of HD video watched is 1,317.9/170.7~J=7.72~J, while we calculate the same value for the case of HEVC video encoding: 2,635.5/341.3~J=7.72~J. Later on, we will discuss the burst mode as a non-functional technology in which the HEVC case could reduce the energy consumption by a factor of 4 even in absence of the non-functional technologies of Section \ref{sec_Practices_Best_Practices_NonFunctional_Technologies_Access_Network}.} This in contradiction with the conclusion of Section \ref{sec_100_vs_2_acess} that stated enhancements have little impact in the case of 40~Gbps access technologies. In the next section, a approach to pricing of resource consumption toward promoting use of best practices, and in particular non-functional technologies, is presented.

\begin{table*}[!htb]
\tiny
\centering
\setlength{\tabcolsep}{2.5pt}
\begin{tabular}{||l|l||c|c||c|c||c|c||c|c||c|c||}\hline\hline
\multicolumn{2}{||c||}{} & \multicolumn{10}{c||}{Technologies}\\\hline\hline
Scenario & Encoding & \multicolumn{2}{c||}{Functional Ta} & \multicolumn{2}{c||}{Functional Tb} & \multicolumn{2}{c||}{Functional Tc} & \multicolumn{2}{c||}{Functional Td} & \multicolumn{2}{c||}{Functional Te} \\\hline
\multicolumn{2}{||c||}{} & \begin{tabular}{c} w/o\\ non-func.\end{tabular} & \begin{tabular}{c} w/\\ non-func.\end{tabular}  & \begin{tabular}{c} w/o\\ non-func.\end{tabular} & \begin{tabular}{c} w/\\ non-func.\end{tabular}  & \begin{tabular}{c} w/o\\ non-func.\end{tabular} & \begin{tabular}{c} w/\\ non-func.\end{tabular}  & \begin{tabular}{c} w/o\\ non-func.\end{tabular} & \begin{tabular}{c} w/\\ non-func.\end{tabular}  & \begin{tabular}{c} w/o\\ non-func.\end{tabular} & \begin{tabular}{c} w/\\ non-func.\end{tabular}  \\\hline\hline
\multirow{2}{*}{Sc1} & AVC & 
\nonecolor\begin{tabular}{c}$\text{BW}_{\text{D}}=6 \text{Mbps}$ \\ \textcolor{BrickRed}{\xmark}
 \end{tabular} & \nonfunc\begin{tabular}{c}$\text{BW}_{\text{D}}=6 \text{Mbps}$ \\ \textcolor{ForestGreen}{\cmark} \end{tabular}&
\nonecolor\textcolor{BrickRed}{\xmark} & \nonfunc\textcolor{BrickRed}{\xmark}&
\nonecolor\textcolor{BrickRed}{\xmark} & \nonfunc\textcolor{BrickRed}{\xmark}&
\nonecolor\textcolor{BrickRed}{\xmark} & \nonfunc\textcolor{BrickRed}{\xmark}&
\nonecolor\textcolor{BrickRed}{\xmark} & \nonfunc\textcolor{BrickRed}{\xmark} \\\cline{2-12}
								 & HEVC & 
\hevc\begin{tabular}{c}$\text{BW}_{\text{D}}=3 \text{Mbps}$ \\ \textcolor{BrickRed}{\xmark} \end{tabular}&  \nonfunchevc\begin{tabular}{c}$\text{BW}_{\text{D}}=3 \text{Mbps}$ \\ \textcolor{ForestGreen}{\cmark} \end{tabular}&
\hevc\textcolor{BrickRed}{\xmark} &  \nonfunchevc\textcolor{BrickRed}{\xmark}&
\hevc\textcolor{BrickRed}{\xmark} &  \nonfunchevc\textcolor{BrickRed}{\xmark}&
\hevc\textcolor{BrickRed}{\xmark} & \nonfunchevc\textcolor{BrickRed}{\xmark}&
\hevc\textcolor{BrickRed}{\xmark} &  \nonfunchevc\textcolor{BrickRed}{\xmark} \\\hline\hline

\multirow{2}{*}{Sc2} & AVC & 
\nonecolor\begin{tabular}{c}$\text{BW}_{\text{D}}=12.1 \text{Mbps}$ \\ \textcolor{BrickRed}{\xmark} \end{tabular} & \nonfunc\begin{tabular}{c}$\text{BW}_{\text{D}}=12.1 \text{Mbps}$ \\ \textcolor{BrickRed}{\xmark} \end{tabular} &
\nonecolor\textcolor{BrickRed}{\xmark} & \nonfunc\textcolor{BrickRed}{\xmark}&
\nonecolor 1,209.0 & \nonfunc 1,209.0 &
\nonecolor 1,119.6 & \nonfunc 1,119.6 &
\nonecolor 1,770.2 & \nonfunc 1,770.2 \\\cline{2-12}
								 & HEVC & 
\hevc\begin{tabular}{c}$\text{BW}_{\text{D}}=6.55 \text{Mbps}$ \\ \textcolor{ForestGreen}{\cmark} \end{tabular} &  \nonfunchevc\begin{tabular}{c}$\text{BW}_{\text{D}}=6.55 \text{Mbps}$ \\ \textcolor{ForestGreen}{\cmark} \end{tabular} &
\hevc 1,410.7 & \nonfunchevc 1,410.7 &
\hevc 2,233.2 &  \nonfunchevc 2,233.2  &
\hevc 2,067.1 &  \nonfunchevc 2,067.1 &
\hevc 3,266.5 &  \nonfunchevc 3,266.5 \\\hline\hline

\multirow{2}{*}{Sc3} & AVC & 
\nonecolor\begin{tabular}{c}$\text{BW}_{\text{D}}=11.1 \text{Mbps}$ \\ \textcolor{BrickRed}{\xmark} \end{tabular} & \nonfunc\begin{tabular}{c}$\text{BW}_{\text{D}}=11.1 \text{Mbps}$ \\ \textcolor{BrickRed}{\xmark} \end{tabular} &
\nonecolor\textcolor{BrickRed}{\xmark} & \nonfunc 390.22 &
\nonecolor 1,317.9 & \nonfunc 617.8 &
\nonecolor 1,220.3 & \nonfunc 572.0 &
\nonecolor 1929.3 & \nonfunc 904.3 \\\cline{2-12}
								 & HEVC & 
\hevc\begin{tabular}{c}$\text{BW}_{\text{D}}=5.55 \text{Mbps}$ \\ \textcolor{ForestGreen}{\cmark} \end{tabular} &  \nonfunchevc\begin{tabular}{c}$\text{BW}_{\text{D}}=5.55 \text{Mbps}$ \\ \textcolor{ForestGreen}{\cmark} \end{tabular} &
\hevc 1,664.9 &  \nonfunchevc 780.4 &
\hevc 2,635.5 &  \nonfunchevc 1,235.4 &
\hevc 2,439.3 & \nonfunchevc 1,143.4 &
\hevc 3,854.3 & \nonfunchevc 1,806.7
\\\hline\hline

\multirow{2}{*}{Sc4} & AVC & 
\nonecolor\begin{tabular}{c}$\text{BW}_{\text{D}}=29.6 \text{Mbps}$ \\ \textcolor{BrickRed}{\xmark} \end{tabular} & \nonfunc\begin{tabular}{c}$\text{BW}_{\text{D}}=29.6 \text{Mbps}$ \\ \textcolor{BrickRed}{\xmark} \end{tabular} &
\nonecolor\textcolor{BrickRed}{\xmark} & \nonfunc\textcolor{BrickRed}{\xmark}&
\nonecolor 494.5 & \nonfunc 231.8 &
\nonecolor 458.4 & \nonfunc 214.9 &
\nonecolor 726.1 & \nonfunc 340.4 \\\cline{2-12}
								 & HEVC & 
\hevc\begin{tabular}{c}$\text{BW}_{\text{D}}=14.8 \text{Mbps}$ \\ \textcolor{BrickRed}{\xmark} \end{tabular} &  \nonfunchevc\begin{tabular}{c}$\text{BW}_{\text{D}}=14.8 \text{Mbps}$ \\ \textcolor{BrickRed}{\xmark} \end{tabular} &
\hevc\textcolor{BrickRed}{\xmark} &  \nonfunchevc 435.0 &
\hevc 988.5 & \nonfunchevc 463.4 &
\hevc 915.6 &  \nonfunchevc 429.2 &
\hevc 1,448.0 &  \nonfunchevc 678.8 \\\hline\hline
\end{tabular}
\caption{The energy consumption (in J) of various combinations of technologies and scenarios of Table \ref{tab_scenario_vs_technology_1} per 1~Gb of download. The impact of the non-functional technologies is taken into account as a ratio where applicable. Also, for the technologies Tb and Tc, a split ratio of $S=128$ is considered. The values associated to AVC and HEVC should divided by 170.7 and 341.3 respectively when consumption per 1~second of HD video watched is required instead of consumption per 1~Gb of download. For 4K videos, the denominators are 64 and 128, respectively. {\em Notes}: $1~J = 1~Ws=0.000278~Wh$, $4184~J = 1~\text{kcal}_{\text{th}}$, and $1055.06~J = 1~\text{BTU}$, where $\text{kcal}_{\text{th}}$ and $\text{BTU}$ stand for the thermochemical kilocalories and the British thermal unit, respectively.}
\label{tab_scenario_vs_technology_LCA1}
\end{table*}

It is worth mentioning that the HEVC case has another joint non-functional technology possibility in the form of burst or shot modes of data transmission. In these modes, a considerably higher bandwidth compared to that bandwidth required for the HEVC operation is allocated to the video stream but in a noncontinuous way. This approach has a considerable potential in reduction of the energy consumption and the footprint, especially  in the case of 40~Gbps access technologies. In its naive form, let us assume that the new bandwidth is the bandwidth of AVC (double the bandwidth required for HEVC according to Table \ref{tab_AVC_vs_HEVC_1}). Then, the energy consumption per 1~second of video watches associated would reduced from 2,635.5/341.3~J (7.72~J) to 660.0/341.3~J (1.93~J) in the case of {\em (functional Tb, w/o non-functional technologies, Sc3, HD, HEVC)} tuple. The same combination with AVC, which would not allow a burst-mode transmission because of bandwidth cap, would have a consumption of 1,317.9/170.7~J (7.72~J) which is 4 (=7.72/1.93) times higher compared to the burst-mode HEVC case. It is worth adding that we have assumed that the ONUs go into a `quasi-off' mode during the intervals of no transmission between the bursts.

\section{Role of Policies and Control in Enforcing Utilization of Best Practice}
\label{sec_Role_Policies_Control_Enforcing_Utilization_Best_Practice}
In comparison of the results in Section \ref{sec_100_vs_2_acess} and \ref{sec_LCA_footprint_Comparison_100_and_2_Cases}, it has been observed that many enhancements and non-functional technologies, which had been before seen essential in delivering bandwidth-hungry services, will be negligible when new 40~Gpbs access technologies come into the picture. Despite this fact, these enhancements could highly improve the efficiency of the operation in terms of energy consumption, and therefore reduce the environmental footprint of the broadband services, especially those related to OVSs. For example, H.265/HEVC which the best practice of video encoding could be mentioned. The operators may continue use the H.264/AVC encoders simply because of availability of enough and extra bandwidth. Although it might be argued that the operators and providers would have a self-drive to switch to more efficient technologies in order to provide a better user experience, the role of legacy hardware and software installation of an operation should not be ignored. The associated cost and management bandwidth required for such a switching move may nullify the self-drive, and therefore external enforcement and drive would be necessary. Also, it is worth mentioning that the notion of best practice used in this section is not limited to video encoding, and the non-functional technologies listed but not limited to in Section \ref{sec_Practices_Best_Practices_NonFunctional_Technologies_Access_Network} could be mentioned among others.

It seems that the current situation, which lacks a sustainability-friendly vision in management of the best practices and their associated pricing, has resulted in a considerable amount of consumption and footprint that could be otherwise avoided. Many operators might have ignored implementation of best practices without being differentiated from those who followed those practices. We propose to handle such a topic by implementing i) differential pricing of resource consumption based on the how much the operation is `far behind' the best practice and ii) also flexible and `inclusive'\footnote{The right to include has been widely considered as a voluntarily approach to sustainable future \cite{Farrahi2014e}.} pricing of the best practices. Although it seems that the second requirement is somehow difficult to be approached, hybrid mechanisms such as those which allow sharing of the differential revenue of the requirement (i) with the holders of the best practices in the form of overlapped fees, could lead to cancellation of charges and reduced total cost while motivating both operator's and best practice holder's actions toward more sustainable future.

The first approach to promote use of non-functional technologies, when they could no longer be considered as temporary functional requirements, is to set a skewed price for resource consumption. In other words, any additional consumption of other practices relative to the best practice is charged at a higher rate. We call this model the Best Practice Delta Factor (BPDF).

The proposed BPDF model can be described as follows. If an operator who uses the practice A requires  $E_A$~kWh ($>$1~kWh) of electricity to deliver a service request while delivering the same service using the best practice B would consume 1~kWh electricity, then the difference in energy consumption and the new price per 1~kWh is defined as follows:
\begin{eqnarray}
\Delta_{\text{A} \rightarrow \text{B}} & = & \left(E_\text{A} - 1\right) \\
c_\text{elec, 1\text{kWh}}^{\text{BPD}} & = & K \Delta_{\text{A} \rightarrow \text{B}} c_\text{elec, 1\text{kWh}} 
\end{eqnarray}
where $c_\text{elec, 1kWh}$ is the price of electricity in dollars per 1~kWh, and $K$ is a factor that determines the intensity of the model. Therefore, the total energy cost for a delivering the service using the practice A would be (while the service cost imposed to the end user is assumed to be \$1 from all providers, i.e, the same cost for service in a competitive environment):
\begin{equation}
C = 1\text{kWh} c_\text{elec, 1\text{kWh}} + \Delta_{\text{A} \rightarrow \text{B}} c_\text{elec, 1\text{kWh}}^{\text{BPD}},
\end{equation}
which shows a nonlinear second-degree dependency on the difference in energy consumption, and therefore it is more sensitive to even small efforts to reduce the consumption.

However, in practice, the best practices are not usually fee-free. Therefore, the BPDF model is updated to take into account the licensing fee of the best practice. In the updated model, the difference in the energy consumption is virtually reduced in order to take into account the impact of the licensing fee:
\begin{eqnarray}
\Delta_{\text{BP}, \text{A} \rightarrow \text{B}} &= & \left(E_\text{A} - 1\right) / (1 + K f_{\text{BP}, \text{B} \rightarrow \text{A}}/\$1) \\
c_\text{elec, 1\text{kWh}}^{\text{BPD}} & = & K \Delta_{\text{BP}, \text{A} \rightarrow \text{B}} c_\text{elec, 1\text{kWh}} 
\end{eqnarray}
where $f_{\text{BP}, \text{B} \rightarrow \text{A}}$ is the license fee per service for a provider, whose baseline the practice is A, to use the best practice of B ($f_{\text{BP}, \text{B} \rightarrow \text{A}}<$ \$1). If a provider that uses the best practice B additionally imposes an increased cost for service,\footnote{Probably because the best practice would also result in better quality of service or quality of experience.} the variable $f_{\text{BP}, \text{B} \rightarrow \text{A}}$ should also include that price difference. The updated cost of energy consumption (resource consumption) would then be:
\begin{eqnarray}
C & = & 1\text{kWh} c_\text{elec, 1\text{kWh}} + \Delta_{\text{A} \rightarrow \text{B}} c_\text{elec, 1\text{kWh}}^{\text{BPD}}, \nonumber \\
 & = & \left( 1\text{kWh} + K \frac{\Delta_{\text{A} \rightarrow \text{B}}^2}{1 + K f_{\text{BP}, \text{B} \rightarrow \text{A}}/\$1} \right) c_\text{elec, 1\text{kWh}}.
\end{eqnarray}

In the BPDF model, it is assumed that if the holder of a best practice waive the associated licensing fees, they would be considered as shareholders of the differential revenue generated by the increased resource consumption price rates. The approach would provide dynamic and competitive means to shift the operations toward more sustainable states considering a selfish nature for all involved actors (the operator and the holder of the best practice, among others). From the perspective of the holder of the best practice, the net revenue from the best practice could be formulated as an optimization problem (for example, in a revenue-per-service-request model):
\begin{eqnarray}
\widehat{f}_{\text{BP}, \text{B} \rightarrow \text{A}} & = & \argmax_{f_{\text{BP}, \text{B} \rightarrow \text{A}}} J\left[f_{\text{BP}, \text{B} \rightarrow \text{A}}\right], \nonumber\\
J\left[f_{\text{BP}, \text{B} \rightarrow \text{A}}\right] & = &
f_{\text{BP}, \text{B} \rightarrow \text{A}} + \nonumber\\
& &  \left(K \frac{\Delta_{\text{A} \rightarrow \text{B}}^2}{1 + K f_{\text{BP}, \text{B} \rightarrow \text{A}}/\$1} \right) c_\text{elec, 1\text{kWh}}.
\end{eqnarray}
In this equation, it is assumed that all revenue generated by the BPDF model is transferred to the holder of the best practice.

\section{Zero Footprint is the new Footprint Reduction: Beyond `Additionality'}
\label{sec_Zero_new_Reduction_Beyond_Additionality}
We would like to briefly mention a side point that can be concluded from the results of the previous sections. The difference in energy consumption after implementing various options has been negative in Table \ref{tab_scenario_vs_technology_LCA1}. In other words, many options, ranging from change in the access technology, to use of non-functional technologies, and to use of video encoding enhancements, could result in reduced energy consumption, and therefore in `footprint reduction' of an operation. However, in many cases, there is one option whose footprint reduction is more than the other options. More precisely, having a reduced footprint with respect to a baseline does not seem to be `conclusive' enough to promote an option. More thorough analysis of other options is required in order to rule out possible misleading and inefficient options that at first may look promising. One way to approach such a challenge is to set the goal to target `zero footprint' instead of footprint reduction. It is worth mentioning that this argument could be imagined as a generalization to the `additionality' requirement. This requirement, which is a well-accepted practice in footprint assessment \cite{ISO140642}, requires that a `green' project satisfies to be additional before proceeding to assign any credit to the reduced emissions of that project with respect to a Business as Usual (BAU) that could later be used to offset (possibly for a fee) the footprint of another operation \cite{Schneider2007}. The logic behind the additionality requirement has been differentiating footprint reduction actions that are business-driven from those that are environmentally-friendly In short, a project that satisfies the additionality requirement would not be implemented without considering the (promised) offset credits in its business model. 

We argue that even the additionality requirement would not be sufficient in future because the scale of technological change and disruption would be beyond any assumed BAU. In such a landscape, fee-for-footprint penalty measures would be more preferable compared to cap-and-trade ones. The penalty obviously becomes zero when a zero-footprint operation is achieved.

\section{Conclusion}
A comparative study of the 40~Gbps metro access technologies and their impact on the broadband access services has been performed. To be precise, various scenarios along with different video encoding technologies and split ratios are considered. It has been observed that these new functional technologies could simply make many of innovative but non-functional technologies, which were before essential to deliver the services, impractical. Therefore, a considerable amount of avoidable consumption and footprint would be expected to be generated. In other words, although with the new access technologies the footprint will be reduced compared to their old access counterparts, this footprint reduction would not be at its optimal and maximal possible level if supporting non-functional technologies are not considered in combination of the main functional technologies. As a direct consequence of this fact, the baseline of any footprint reduction project would be highly ill-defined and subjective, and therefore all performance measures of footprint reduction that are based on a baseline would be questionable. As a starting point to address this dilemma, a new, smart approach to pricing of resource consumption, especially electricity consumption, has been proposed in order to make the non-functional technologies applicable, and also promote utilization of the best practices in the operations.

\section*{Acknowledgment}
The authors thank the NSERC of Canada for their financial support under Grant CRDPJ 424371-11 and also under the Canada Research Chair in Sustainable Smart Eco-Cloud (NSERC-950-229052).

\begingroup
\bibliographystyle{IEEEtran} \bibliography{imagep}
\endgroup

\setcounter{section}{0}
\renewcommand{\thesection}{Appendix \arabic{section}}
\section{LCA for ICT: Adaptation Challenges}
\label{sec_Challenges_Adaptation_LCA_ICT}
Life Cycle Assessment (LCA) for ICT has been correctly identified as an obligatory mandate by many researchers and operators \cite{Arushanyan2014,Farrahi2014c}. Many efforts have been put in place toward adapting already-in-use LCA approaches to the ICT solution. In this appendix, we attempt to highlight a few aspects of the LCA for ICT that would pose serious challenges for the methodology developed to assess the life cycle impacts of the ICT solutions.

Life cycle assessment have been the flag ship in the assessment of the environmental impact of products, especially in the form of extending the scope by: i) considering upstream (manufacturing and related process), downstream (end-of-life processes), and sidestream (indirect influence on the external processes) phases in addition to the use phase and ii) including other impacts than Carbon Dioxide ($CO_2$) and Green House Gases (GHG) emissions. Various organizations have weighed in developing LCA standards and guidelines, such as ISO (14040:2006, 14044:2006, 14071:2014), ETSI (EU) TS 103 199 (V1.1.1):2011, WRI, and WBCSD. Although application of LCA has been highly successful to various industries, many challenges have been identified that are mainly exclusive to ICT industry \cite{ECOFYS2013}. 

More fundamentally, in contrast to many industries, in the case of the ICT industry, the role of a product or a service `instance' has been disappearing and immersing in the whole `experience' between a user and their provider. In other words, assessment of a product or a service would probably give a partial picture, and in order to build a complete one, assessment of an ICT technology should be directly considered.

\subsection{Challenge of Life Time Definition}
In contrast to many products, the `calendar' age of an ICT product would not exactly represent its true age. Although, because of possibly some ``out-of-phase'' phenomena, the efficiency of a product would not be the same as that of a similar product within its designated calendar life, the product would be still usable probably for a different mission and in a different location till finishing its nominal working hours. A simple year-based assessment, such as that of the typical 5-year life time assumption,  would be prune to error and bias, and it might also result in preventing a potential technology, which would otherwise make use of the functioning but outdated ICT products and `services' for a better good, to get public approval and support. A good example is the Rainforest Connection (RFCx) Project that uses outdated cell phones and the GSM technology \cite{Gross2014,Ferguson2013} to protect rain forests.

\subsection{Challenge of End-of-Life}
In contrast to many products, the ICT products are not just limited to natural resource extraction. Therefore, their end-of-life requirements cannot be simply fulfilled by returning the e-waste material back to the nature. For example, the health issues related to the `escaped' ICT e-waste especially in the unmonitored regions would require a special attention and management. In other words, the responsibility of an ICT product should be carried forward by its owner indefinitely.

\end{document}